\documentstyle[12pt]{article}
\newcommand{\beq}{\begin{equation}}
\newcommand{\eeq}{\end{equation}}
\newcommand{\beqa}{\begin{eqnarray}}
\newcommand{\eeqa}{\end{eqnarray}}
\newcommand{\bec}{\begin{center}}
\newcommand{\eec}{\end{center}}

\begin{document}
 \title{ Integrability and Computability in Simulating Quantum Systems}
\author{K. Umeno\thanks{E-mail: chaosken@giraffe.riken.go.jp}\\
Laboratory for Information Representation, 
Frontier Research Program\\
The Institute of Physical and Chemical Research (RIKEN)\\ 2-1 Hirosawa,
Wako, Saitama 351-01, Japan}
\date{}
\maketitle 
\begin{abstract}
    
 An impossibility theorem on approximately   
 simulating quantum non-integrable Hamiltonian 
 systems  is presented here.
 This result shows that there is a trade-off between the unitary property and 
  the energy expectation conservation law in time-descretization of 
  quantum non-integrable systems, whose classical counterpart is  
 Ge-Marsden's impossibility result about simulating  
classically non-integrable Hamiltonian 
   systems using integration schemes preserving symplectic (Lie-Poisson) 
  property. 
\end{abstract} 

\section{Introduction}

 Recently, much attentions are directed  to investigate 
 the interrelation between 
physics and computation. 
 To connect physics with computation, we can classify the problems
 into the
  the following classes:\\[0.3cm] 
 class (1): Connection between classical physics and classical computation,\\
 class (2): Connection between quantum physics and classical computation,\\
 class (3): Connection between classical physics and quantum computation, and \\
 class (4): Connection between quantum physics and quantum computation.\\
                                                                          
Concerning the class (4),
 {\it simulating} quantum behavior such as  quantum chaos using classical
computers
  is  known to be
  a notoriously  difficult computational problem\cite{feynman:paper2}. 
One of the main difficulties is that one must discretize a continuous  
time parameter of equations of motion in order to simulate on 
computers.  
Thus, it is an important question to ask whether we can always 
have a suitable time-discretization 
scheme of Schr\"odinger equation.

In this paper, I will give a somewhat 
 negative answer to this question: In the 
case of {\it quantum non-integrable} systems with  an explicit 
time-independent  
 Hamiltonian operator, no {\it explicit} time-discretization  
algorithm preserving unitary property  
  can simulate quantum non-integrable behavior without 
violating the conservation law of energy expectation.
Since the original 
 quantum nature must have these two properties, namely, the conservation law of 
 energy expectation and the unitary property of time evolution,
  this means that there is a 
  fundamental limit in simulating quantum non-integrable behavior using 
  unitary maps like quantum computers.     
  This negative result can be regarded as a quantum analogue of  
 Ge-Marsden's theorem\cite{zhong:paper}:
 No symplectic integrator can simulate 
 non-integrable behavior in a class of  autonomous Hamiltonian systems without
 violating the energy conservation.  
 These  fundamental limits, whether quantum or classical, 
 suggest the importance of the notion of integrability in  
    simulating 
  physical behavior.  In Section 2, we give a brief explanation of 
  time-discretization preserving unitary property. 
In Section 3, we give a theorem about the impossibility of simulating 
quantum non-integrable systems. 
In Section 4, we discuss various aspects about our results. 
\section{Simulation technology preserving unitary property}
Time-evolution of quantum computation 
can be seen as a class of successive iterations of    
 unitary  
transformations\cite{benioff:paper,deutsch:paper1,feynman:paper3}. 
The time-evolution operator has the form of  
\begin{equation}
  U(\Delta t)= \exp\left[-i\Delta t H/\hbar\right],
\end{equation} 
 where \(H\) is a Hamiltonian operator with Hermitian property, \(\Delta t\) is 
 the time duration 
 of each computation process and 
 an exponential operator \(\exp\left[x A\right]\) 
 is defined as  
\begin{equation}
\exp\left[x A\right]=\sum_{n=0}^{\infty}\frac{(xA)^{n}}{n!}, 
\quad x=-i\Delta t/\hbar. 
\end{equation}
Let \(A\) and \(B\) be Hermitian operators as the 
generators of two {\it different} elementary processes of unitary dynamics. 
In general, \(A\) does not commute with \(B\):
\begin{equation} 
 \left[A,B\right]=AB-BA\neq 0.
\end{equation}
To track  computational processes   retaining unitary property, 
evaluating the following time-evolution operator
\begin{equation}
\label{eq:expplus} 
\exp\left[x (A+B)\right]
\end{equation} 
 is  relevant to various problems. In fact, there are 
 infinitely many methods to get perturbation series of Eq.(\ref{eq:expplus}).
The  Feynman path-integral method\cite{feynman:paper1} 
\begin{equation}
  \exp\left[x (A+B)\right] \approx \left[1+\frac{x(A+B)}{n}\right]^{n}
\end{equation}
 discovered in his study of quantum electro-dynamics  
is a first-order method
based on the identity  
\begin{equation}
\exp\left[x (A+B)\right]=\lim_{n\rightarrow \infty} 
          \left[1+\frac{x(A+B)}{n}\right]^{n}. 
\end{equation}

 However, the above approximation breaks unitary property in each 
elementary dynamical process 
 \(1+\frac{x(A+B)}{n}\), as is easily checked.  
 On the contrary,   
Trotter formula \cite{trotter:paper}
\begin{equation}
\label{eq:trotter}
 \exp\left[x (A+B)\right]=
          \left[\exp (\frac{xA}{n}) \exp (\frac{xB}{n}) \right]^{n}
  + \mbox{\cal O} \left(\frac{x^{2}}{n}\right)
\end{equation}
based on the identity 
\begin{equation}
\exp\left[x (A+B)\right]=\lim_{n\rightarrow \infty}
          \left[\exp (\frac{xA}{n}) \exp (\frac{xB}{n}) \right]^{n}
\end{equation}
preserves the unitary property in each elementary process 
\(\exp (\frac{xA}{n}) \exp (\frac{xB}{n})\),  as is also  easily checked.  
The second order formula called Leap flog method has the form 
\begin{equation}
\exp\left[x (A+B)\right]=
          \left[\exp (\frac{xA}{2n}) \exp (\frac{xB}{n}) 
               \exp (\frac{xA}{2n})\right]^{n}
  + \mbox{\cal O} \left(\frac{x^{3}}{n^{2}}\right).
\end{equation}
Furthermore,  
 many other higher-order formulas for exponential 
operators \(\exp\left[x (A+B)\right]\) preserving the symmetry corresponding to the unitary 
propery  were recently 
 discovered  independently both in the development of 
 simulation technology called  quantum Monte Carlo methods 
\cite{suzuki:book1,suzuki-umeno:paper} to simulate 
 density matrices, or in the development  of  simulation technology called 
  symplectic integrators
\cite{ruth:paper,suzuki:paper2,suzuki:paper3,suzuki-umeno:paper,umeno:paper1,yoshida:paper2}
   to simulate classical Hamiltonian dynamical systems. 
It is an easy task to 
  extend these decomposition formula of the exponential operators
\(\exp\left[x(A+B)\right]\) to  more generalized exponential operators
\(\exp\left[x \sum_{j=1}^{l} A_{j}\right]\) of multi noncommutative
operators \(A_{1},A_{2},\cdots, A_{l}\).
 Thus, 
an application of these successive composition formulas 
of exponential operators  
 to quantum computations of  
\(\exp\left[x \sum_{j=1}^{l} A_{j}\right]\) 
can give us a unified view of this kind of simulations as follows:
Let us consider the problem of approximately simulating 
\(\exp\left[x \sum_{j=1}^{l} A_{j}\right]\) for 
 \(t\leq t' \leq t+\Delta t\)  based on an explicit algorithm on 
 quantum model of computation whose each elementary process 
 is successively generated by  explicitly 
  time-dependent Hamiltonians \(Q_{j}(t,\Delta t),1\leq j \leq m \). 
 Then, each \(s\)-th order approximation formula has a form:
\begin{equation}
\label{eq:explicitq}
\exp\left[x \sum_{j=1}^{l} A_{j}\right]=
\prod_{j=1}^{m}\exp\left[x Q_{j}(t,\Delta t)\right]+ \mbox{\cal O} 
(x^{s+1}).
\end{equation} 
The relation 
\begin{equation}
\label{eq:relation}
\sum_{j=1}^{l} A_{j}=\sum_{j=1}^{m} Q_{j}(t,\Delta t)
\end{equation}
  must hold from    
the lowest order terms in \(x\)
 in Eq. (\ref{eq:explicitq}). 
 
\section{Theorem} 
Let us consider a time-independent  
 Hamiltonian \(H(\mbox{\boldmath$q,p$})\)  
 in a certain class 
of the set of Hermitian operators
\(\tilde G= \{G(\mbox{\boldmath$q,p$})\}\), 
where \(\mbox{\boldmath$q$}\) and \(\mbox{\boldmath$p$}\) denotes the 
canonical conjugate operators in the  standard sense of quantum mechanics.   
We can define {\it quantum non-integrability} as follows:\\
\newtheorem{df}{Definition}
\begin{df}
We call a quantum Hamiltonian system with a time-independent Hamiltonian 
operator \(H\) quantum non-integrable if the following relation holds:
\begin{equation}
  \left[\Phi, H \right]=0\Longrightarrow \Phi=  F(H),   
\end{equation} 
 where \(\Phi \in \tilde G \) and \( F\) 
 is a some function 
of a variable. 
\end{df}
 
Since \(H\) is a time-independent Hamiltonian operator, the 
expectation value of \(H\) must be  preserved:
\begin{equation}
\frac{d}{dt}<H>=\frac{d}{dt}<\Psi |H|\Psi>=0,
\end{equation}
 where \(<\Psi |\) is the state vector.
 
Here,  we prove the following theorem:\\

 \newtheorem{theorem}{Theorem }
\begin{theorem}
If an explicit algorithm preserving unitary property  
can simulate a quantum 
non-integrable system with a time-independent Hamiltonian \(H\) approximately,  
 the conservation law of the expectation value of the Hamiltonian 
operator \(<H>\) must break down.
\end{theorem}    
{\bf Remark 1:}\ \
This theorem does {\it not depend} on the order and types of approximate
algorithms  we choose.\\
{\bf Remark 2:}\ \ 
A class of explicit algorithms preserving unitary property involves 
universal quantum Turing machines in the sense of Deutsch\cite{deutsch:paper1}.
Thus, as is shown in Ref. \cite{umeno:paper5,umeno:paper6}, 
 this theorem shows that there is no (discrete time) quantum computers
 to simulate quantum non-integrable 
systems without breaking  
 the conservation law of the energy expectation. However, the 
 present theorem says not only the limitation of quantum computers on this aspect but also
 a more general statement  
 that there is a universal trade-off between the unitary property  and 
 the conservation law of energy expectation in time-discretization of 
 quantum non-integrable systems.\\
 
{\bf (Proof of Theorem 1)}\\
By using   
the expression of quantum algorithms in 
Eq. (\ref{eq:explicitq}), we can consider an   
 \(s\)-th order algorithm of approximately simulating the quantum 
dynamics of \(H\) for the time duration \(\Delta t\) of a computational 
step  as follows:
\begin{equation}
\label{eq:H_explicit}
\exp\left[x H\right]=
\prod_{j=1}^{m}\exp\left[x Q_{j}(t,\Delta t)\right]+ \mbox{\cal O}(x^{s+1}),
\end{equation}
 where \(x=-i\Delta t/ \hbar\) and \(1\leq s < \infty \). 
Each quantum algorithm \(Q_{j}(t,\Delta t)\) 
has a corresponding {\it time-dependent} Hamiltonian 
 \(H_{j}(t)\) satisfying 
\begin{equation}
    Q_{j}(t,\Delta t)= T(\exp \int_{t}^{t+\Delta t} H_{j}(s)ds)=1+\sum_{n=1}^{\infty} 
   (-\frac{i}{\hbar})^{n}\int_{0}^{t_{1}}dt_{1}\cdots \int_{0}^{t_{n-1}}dt_{n}H_{j}(t_{1})\cdots 
   H_{j}(t_{n}),
\end{equation} 
where \(T\) denotes the time ordering. 
The resulting  quantum algorithm has also an {\it time-dependent} Hamiltonian 
\(\tilde H(t,\Delta t)\) satisfying the relation 
\begin{equation}
\label{eq:effective}
\prod_{j=1}^{m}\exp\left[x Q_{j}(\Delta,t)\right]=\exp (x \tilde H(t,\Delta t)).
\end{equation}
 By successively applying the  
Baker-Campbell-Hausdorff formula:
\begin{equation}
 \mbox{exp}X\mbox{exp}Y=\mbox{exp}Z,
\end{equation} 
where
\begin{equation}
 Z=X+Y+\frac{1}{2}[X,Y]+\frac{1}{12}\left([X,[X,Y]]+[Y,[Y,X]]\right)
+\frac{1}{24}[X,[Y,[Y,X]]]+\cdots
\end{equation}
to the system (\ref{eq:effective}), we can compute 
the corresponding time-dependent 
Hamiltonian 
\(\tilde H\) in a form:
\begin{equation}
 \tilde H(\mbox{\boldmath$q,p$},t,\Delta t)= H+\sum_{n=s}^{\infty}(\Delta t)^{s}H_{s}(t)= 
H +\mbox{\cal O} (x^{s}), 
\end{equation}
 where \(H_{s}(t)\) is a time-dependent 
 correction term of order \(s\).
We assume that the energy expectation \(<\tilde H>\) in the quantum 
simulation  is also 
preserved: 
 \begin{equation}
\label{eq:relation_H}
 <H>=<\tilde H>=\mbox{Const.} \quad \mbox{for}\quad 
 t\leq t'\leq t+\Delta t.
\end{equation}
Since we can choose \(\Delta t \) 
an arbitrary real number,
 the relation (\ref{eq:relation_H}) means 
the following commutation relations hold:
\begin{equation}
\label{eq:commutation}
\left[ H, \tilde H(t)\right]=0\quad\mbox{and}\quad 
\left[ H, H_{n}(t)\right]=0 \quad \mbox{for}\quad n\geq s.
\end{equation}
However, 
from
 the assumption of  
 {\it quantum non-integrability} of \(H\), it follows that  
 that \(\tilde H =   F (H) \). This means that 
the quantum algorithm 
  \( \tilde H\) generates the {\it exact} quantum dynamics of \(H\). 
 This exactness \((s \rightarrow \infty )\) 
 contradicts the assumption that 
the underlying quantum algorithm gives an {\it approximate } 
tracking  of the dynamics of \(H\) in the finite 
 order \(s\).\\ 
{\bf (End of Proof)}

\section{Discussions} 
  The key of the present analysis is in {\it quantum non-integrability}. 
  How generic is the notion of quantum non-integrability in quantum 
  mechanics?
  In classical mechanics, it is known that most dynamical systems are 
  non-integrable since the famous Poincar\'e theorem in the last century. 
Furthermore,  we have exact criteria of
    classical non-integrability for  explicitly given  Hamiltonian systems
      based on the singularity analysis
\cite{ito:paper,umeno:paper2,umeno:paper3,umeno:paper4,yoshida:paper1,ziglin:paper1,ziglin:paper2}.
  
  On the contrary, in quantum mechanics,  
  we do not have any theorem guaranteeing the generic 
    character of quantum non-integrable systems corresponding to the Poincar\'e 
   theorem in classical mechanics nor   
      exact criteria of quantum non-integrability for  explicitly given 
      Hamiltonian operators.
   In other words, it is  
      not a trivial thing to connect 
      classical non-integrability with quantum non-integrability 
      \cite{hietarinta:paper,weigert:paper}. Recently, the present author 
      found that the quantum Hamiltonian system with a time-independent 
     Hamiltonian operator
     \(H=\frac{1}{2}(p^{2}_{x}+p^{2}_{y}+q_{1}^{2}q_{2}^{2})\)
         would be
     quantum-nonintegrable under the hypothesis of the Weyl rule for 
     canonical variables \(p_{i},q_{i}\) using the Moyal bracket, based on 
     Ziglin's result of proof of its classical non-integrability  
     \cite{umeno:paper6}. For the classical system of this system, 
    it was shown in Ref. 
\cite{umeno:paper1} that we cannot avoid energy fluctuations for some specific  
     initial conditions like \((q_{1},q_{2},p_{1},p_{2}=(1000,0.002,0,0))\)  
    because the higher-order correction terms \(H_{s}\) also become  
    bigger as 
\begin{equation}
         |H_{s}| \approx AB^{s},
\end{equation}   
   where \(A\) and \(B\) are some positive real constants. It can be easily 
  predicted that  
 this divergence of the higher-order correction terms  \(H_{s}\) 
  can also  occur in quantum non-integrable systems like a quantum version of 
  the above system. This model can be a vivid example causing  
  rather general phenomena of the breakdown of 
  the conservation law of energy expectation for quantum non-integrable 
 systems by using any finite-order time-descretization preserving 
 unitary property, which Theorem 1 asserts.    
 This result has an interesting implication concerning 
 the usual  energy-time uncertain relations. 
From the energy-time uncertain relations \(\Delta t\cdot \Delta E\geq \hbar\),
it follows that 
\begin{equation}
\Delta t^{s} |H_{s}| \approx  \Delta E \geq \frac{\hbar}{\Delta t}.
\end{equation} 
This inequality gives a lower-bound of \(\Delta t\) which depends only on  
the order \(s\) of time-discretization and the correction terms \(H_{s}\). 
Thus, this analysis shows that  our naive view that
  the continuous nature of time in quantum mechanics is naturally 
  obtained in the 
  continuum limit \(\Delta t\rightarrow 0 \) is not universal, 
  at least in quantum 
  non-integrable systems.  
 It will be an interesting and important open 
 problems to consider time-discretization 
   of quantum non-integrable systems in connection with the foundation of 
  quantum mechanics.\\   

{\bf Acknowledgements}\\

 This work was supported in part by 
the Special Researcher's Program to promote basic sciences 
   at  RIKEN
 and from the Frontier Research  Program.
 I would like to thank Prof. Shun-ichi
Amari for his
continual encouragement.
  
\end{document}